\documentstyle[aps,psfig]{revtex}  

\begin{document}

\title{Proximity-induced transport
in  hybrid mesoscopic normal-superconducting  metal structures. }

\author{A.F. Volkov$^{+*}$, V.V. Pavlovskii$^{++}$ and R. Seviour$^{*}$}
\address{
* School of Physics and Chemistry,
Lancaster University, Lancaster LA1 4YB, U.K.
\\ + Institute of Radioengineering and Electronics of the Russian
   Academy of Sciencies, Mokhovaya str.11, Moscow
   117218, Russia.
\\ ++ Institute of Physics and Technology of the Russian
   Academy of Sciencies,Krasikov 25a, Moscow
   103907, Russia.}
\date{\today}
\maketitle

\begin{abstract}
 Using an approach based on quasiclassical Green's functions
we present a theoretical study of transport in mesoscopic 
S/N structures in the diffusive limit. 
The subgap conductance in S/N structures with 
barriers (zero bias and finite bias anomalies) are discused.  We also 
analyse the temperature dependence of the conductance variation $\delta S(T)$ 
for a
Andreev interferometer. We show that besides the well know low temperature 
maximum a second maximum near $T_c$ may appear. We present the results of studies on 
the Josephson effect in 4 terminal S/N/S contacts and on the possible 
 sign reversal of the Josephson critical current.
\end{abstract}

\smallskip
 {\bf 1. Introduction}
\smallskip

 The theory describing non-equilibrium transport in superconductors 
was developed 20 years ago (see, for example \cite{r1}) and has been 
used to explain phenomena such as viscous flux flow, Josephson 
effects in superconducting weak links and the passage of the current over 
the interface between a superconductor and a normal 
metal (an S/N interface). Technological advances achieved in 
the last decade have enabled the fabrication of mesoscopic structures 
where the dimensions are less than the energy relaxation length and with 
well defined physical properties. In mesoscopic systems phase 
coherence is maintained and the distribution of quasiparticles may differ 
significantlty from  equilibrium. Experiments on these 
structures have revealed a number of new effects such as, 
subgap conductance in SIN junctions (where I is an insulating layer), 
the oscillatory dependence of the conductance in S/N structures containing 
normal metal or superconducting loops, the non-monotonic dependence of the 
N film conductance on the temperature and voltage in S/N structures; long-range, 
phase-coherent effects, change of sign of the Josephson critical current in 
4-terminal S/N/S structures when an additional current is driven through the N film. 
Although it is only recently that most of these effects have been explained using a 
non-equilibrium  theory developed at the end of the 1970's (see \cite{r2,r3}). 
One method for studying these effects (ballistic systems in particular) 
is based on the Bogolyubov-de Gennes equations and the Landauer formula for
the conductances \cite{r2,r3}. In this paper we use another approach based on 
a microscopic, quasiclassical Green's function technique, which has been 
successfully applied to systems with a small mean free path (i.e. diffusive regime), 
to analyse these effects. The method of matrix, quasiclassical Green's functions 
is a convenient and powerful method for studying transport in mesoscopic S/N structures 
\cite{r4}. The quasiclassical approximation means that all the Green's functions 
are spatially averaged over distances of order of the Fermi wave length $p_F^{-1}$. 
This is  equivalent to the intergration in momentum space over the variable 
$\xi_p=(p-p_F) v_F$. To describe transport and non-equilibrium properties we need 
to know the retarded (advanced) Green's functions $G^{R(A)}$ and the Keldysh function $G$. 
In the case of superconducting systems each of these functions is a 2x2 matrix elements 
of which describe normal excitations and a condensate. These matrices are grouped 
together, defining the 4x4 matrix 
Green's function $\check{G}$ as
\begin{equation}
\check{G} = \left(\begin{array}{cc}
\hat{G}^R & \hat{G} \\ 0 & \hat{G}^A\end{array}\right)
\label{eq1}
\end{equation}
 where  $\hat{G}^{R(A)}$ are the matrix retarded (advanced) Green's functions 
and $\hat{G}$ is the Keldysh function. The functions $\hat{G}^{R(A)}$ 
describe thermodynamical properties of the system such as the density of states 
(DOS), condensate currents and so on, while the function  $\hat{G}$  
describes the kinetic and nonequilbrium properties of the system.
 The matrix $\hat{G}$ is related to a matrix 
distribution function $\hat{f}$ by
\begin{equation}
\hat{G}=\hat{G}^{R} \hat{f} - \hat{f} \hat{G}^{A}
 \label{eq2}
\end{equation}
 Where the elements of the matrix $\hat{f}=f_{o}\hat1+f\hat{\sigma}_{z}$  
 describe particle and hole-like excitations. The component 
 $f_{o}$ is used in calculating the order parameter and  
supercurrent, while $f$ describes the branch imbalance and determines the 
spatial distribution of the electric field (see for example \cite{r4,r5}). 
The matrices $\hat{G}^{{R}(A)}$ can be presented in the form,
\begin{equation}
\hat{G}^{R(A)}=G^{R(A)}\hat{\sigma_{z}} + \hat{F}^{R(A)}
 \label{eq3}
\end{equation}
In a bulk superconductor $G^{R(A)}=\epsilon / \xi_{\epsilon}^{R(A)}$, 
$\hat{F}^{R(A)}=F^{R(A)}\left(i\hat{\sigma}_{x}\sin\chi + i \hat{\sigma}_{y} 
\cos\chi\right),F^{R(A)}=\Delta/\xi_{\epsilon}^{R(A)}$, $\xi_{\epsilon}^{R(A)} 
=\sqrt{(\epsilon \pm i \Gamma)^2-\Delta^2}$, $\Gamma$ is the damping rate in 
the excitation spectrum of the superconductor and $\chi$ is the phase 
of the order parameter.The current in the system is expressed in 
terms of the function $\hat{G}$ \cite{r4,r5},
\begin{equation}
I=(\sigma d/8)Tr\hat{\sigma}_{z}\int d\epsilon 
(\hat{G}^{R}\partial_x\hat{G}+\hat{G}\partial_x\hat{G}^{A})
 \label{eq4}
\end{equation}
 This expression is written for the one-dimentional case, where d is the 
thickness of the system. In the diffusive limit the supermatrix $\check{G}$  
in the N film of a S/N mesoscopic system obeys the equation \cite{r4},
\begin{equation}
 D\partial_{x}\left[\check{G}\partial_{x}\check{G}\right]
 +i\epsilon\left[\check{\sigma}_{z},\check{G}\right]= 0.
 \label{eq5}
\end{equation}
 where $D$ is the diffusion coefficient and $\check{\sigma}_{z}$ is the Pauli 
supermatrix. We neglect the inelastic collision integral by assuming 
that the length of the N film L is short 
enough:$\epsilon_{Th} >>[\hbar \tau^{-1}_{e-e},\hbar \tau^{-1}_{e-ph}]$,  
($\epsilon_{Th}=\hbar D/L^2$ is the Thouless energy and $\tau_{e-e},
\tau_{e-ph}$ are the electron-electron and electron-phonon collision 
times). To solve Eq.(\ref{eq5}) we must have the complementary 
 boundary condition,
\begin{equation}
 D\left(\check{G}\partial_{z}\check{G}\right)=\left(\epsilon
_{b} d_{N} \right) \left[ \check{G} ,\check{G} _{S} \right] 
\label{eq6}
\end{equation}
 the z-axis is normal to the plane of the S/N interface.
The boundary conditions for the quasiclassical Green's functions
 $\check{G}$ were derived for the general case by Zaitsev \cite{r6} 
and reduced to the simple form (\ref{eq6}) by Kupriyanov and Lukichev 
\cite{r7} in the dirty case. The applicability of Eq.(\ref{eq6}) is 
discussed in Ref.\cite{r8}. The distribution functions in the 
reservoirs  $f_0$ and $f$ are assumed to be in equilibrium and for example,
for the structure shown in Fig.1(b) they have the form,
\begin{eqnarray}
f_{o}(\pm L) = F_{oN} (\epsilon) \equiv [\tanh ((\epsilon+eV)\beta)+
\tanh ((\epsilon-eV)\beta)]/2,
\label{eq7}\\
f(\pm L)=  F_{N}(\epsilon) \equiv
[\tanh ((\epsilon+eV)\beta)-\tanh ((\epsilon-eV)\beta)]/2,
 \label{eq8}
\end{eqnarray}
where $\beta=(2T)^{-1}$ and $\pm V$ are the electric potentials in the 
reservoirs at $x=\pm L$ (we have assumed the system is symmetrical).
 In order to find the distribution functions $f(\epsilon,x)$ in the 
N film, we consider element $(12)$ of Eq.(\ref{eq6}) (i.e. we consider 
the equation for the Keldysh matrix $\hat{G}$). Consider for simplicity 
a strucutre of the type in Fig.1(b) with one superconducting strip in 
the centre of the N film. Multiplying this equation by 
$\hat{\sigma}_z$ and taking the trace, we obtain,
\begin{equation}
 \partial_{x}(M(x)\partial_{x}f)=0
 \label{eq9}
\end{equation}
where $M(x)=(1/2)(1-\hat{G}^{R}\hat{G}^{A}-\hat{F}^{R}\hat{F}^{A})$ and we 
have assumed that the width of the S strip ($w$) is small compared to $L$. 
 In this limit  
the distribution functions $f(\epsilon,0)$ and $f_S(\epsilon,0)$ are equal 
to zero ($V(0)=0$). Integrating Eq.(\ref{eq9}) and taking into account 
 boundary condition (\ref{eq8}), we obtain,
\begin{equation}
 J=(F_N/L)<M^{-1}>,
 \label{eq10}
\end{equation}
 here $<M^{-1}>=L^{-1} \int^L_0 dx M^{-1}$ (see \cite{r9}) and $J(\epsilon)$ 
is the integration constant which determines the current,
\begin{equation}
I=(\sigma d/2) \int _0^{\infty} d\epsilon J(\epsilon).
 \label{eq11}
\end{equation}
 Therefore, the problem of finding the $I(V)$ characteristics is reduced to 
 finding the retarded (advanced) Green's functions which 
determine the function $M(\epsilon,x)$. Using the well known substitution, 
$G^{R(A)}=\cosh u^{R(A)}$ and 
$F^{R(A)}=\sinh u^{R(A)}$, the equation for the retarded 
(advanced) Green's functions $\hat{G}^{R(A)}$ may be presented in the form,
\begin{equation}
\partial_{xx} u^{R(A)} - \left(k^{R(A)}\right)^2 \sinh u^{R(A)}=
-k_{b}^{2} w (\delta\left(x+L_{1} \right)+\delta\left(x-L_{1} \right)
)\left[F_{S}^{R(A)}
\cosh u - G_S^{R(A)} \sinh u \right]
\label{eq12}
\end{equation}
 We have used Eqs.(\ref{eq5}) and (\ref{eq6}) and assumed that $w<<L$, where $k_{b}^{2}
=\epsilon_b/D$,$(k^{R(A)})^2=\gamma \mp 2i\epsilon$. The term on the right hand side 
of Eq.(\ref{eq12}) is the source of the condensate (the proximity effect).
 Solutions to Eq.(\ref{eq12}) can be written in an explicit form in 
some limiting cases. For example in the case of a weak proximity effect 
($|F^{R(A)}|<<1$) Eq.(\ref{eq12}) 
can be linearised (i.e. $\sinh u^R \approx  u^R$), then a solution of 
Eq.(\ref{eq12}) is easily  found. This case occurs if the ratio of the 
resistance of the N film $R_N$ and the $S/N$ interface resistance 
$R_{S/N}$ is small, $r \equiv R_N/ R_{S/N}<<1$. 
The comparison of an exact numerical solution and 
the solution of the linearised Eq.(\ref{eq12}) shows that even when 
$r\approx 1$ the difference between these solutions is less than $10\%$.
 Another limiting case corresponds to a short N film ($\epsilon_{Th}>>(T,V)$), 
now Eq.(\ref{eq12}) can be averaged over the length \cite{r10,r11}. We use 
this procedure to analyse the subgap conductance of the structure 
shown in Fig.1(a).

\smallskip
 {\bf 2. Subgap conductance of S/N/N' structures}
\smallskip

 Considering the S/N/N' system shown in Fig.1(a) and assuming the 
middle N film is short, i.e. $\epsilon_{Th} \equiv D/L^2 >> [T,eV]$ ( 
L is the length of the N film and V the applied voltage), we can 
average Eq.(\ref{eq12}) over the length using the boundary condition 
(\ref{eq6}) and obtain for $F^{R(A)}$ and  $G^{R(A)}$,
\begin{equation}
F^{R(A)}=
\frac{\epsilon _{g}}{\sqrt{ \left( \epsilon \pm i\gamma \right)
^2-\epsilon _{g}^2}} 
;G^{R(A)}=
\frac{\epsilon \pm i\gamma }{\sqrt{ \left( \epsilon \pm i\gamma \right)
^2-\epsilon _{g}^2}}
 \label{eq13}
\end{equation}
 Here $\epsilon_{g} = \rho D/(2 R_{bS \Box }L)$,
$\gamma=\rho D/(2 R_{bN \Box }L)$ 
and we have assumed that $\Delta>>(\epsilon,\epsilon_{g})$. $R_{bS,N \Box }$ are 
the interface resistance per unit area. The energy 
$\epsilon_{g}$ is a minigap in the excitation spectrum of the N film 
induced by the proximity effect \cite{r12,r13}. The damping constant $\gamma$ is 
due to the N/N' contact, which increases with decreasing the N/N' interface 
resistance. In Fig.2(a) we show the energy dependence of the DOS in the 
normal film for different $\gamma$,

\begin{equation}
\nu_N=(G^R-G^A)/2
\label{eq14}
\end{equation}

 We see that if $\gamma>>\epsilon_g$ the DOS in the N film for $T<T_c$
is close to the DOS when $T>T_c$, and a subgap arises when $\gamma<<\epsilon_g$. 

 In order to calculate the conductance of the structure, we need to find the 
distribution function $f$ from Eq.(\ref{eq9}), taking into account boundary 
condition (\ref{eq6}). Carrying out some simple calculations, we obtain the 
dimensionless conductance \cite{r9},

\begin{equation}
S= R_N \frac{dI}{dV} =\int _0^{\infty}d\epsilon \beta F_N'
[M^{-1} + \frac{1}{\nu_d} (\frac{\epsilon_{Th}}{\epsilon_g})
(\frac{\epsilon_g}{\gamma})+ \frac{\epsilon_{Th}}{A\epsilon_g}]^{-1}
\label{eq15}
\end{equation}

Here $F_N'=\partial F_N/ \partial (eV)$, M is defined in Eq.(\ref{eq9}) and,

\begin{equation}
A=\frac{1}{4}[(G^R-G^A)(G_S^R-G_S^A)-(F^R+F^A)(F_S^R+F_S^A)]
\label{eq16}
\end{equation}

 The terms in the square brackets in Eq.(\ref{eq15}) determine the normalised 
resistances of the N film, N/N' and N/S interfaces respectively. The first term 
in Eq.(\ref{eq16}) stems from the usual quasiparticle current, which is non-zero 
when $eV>\Delta$. The second term in Eq.(\ref{eq16}) represents the subgap 
conductance due to Andreev reflection processes and is non zero when 
$\epsilon<\Delta$ (if the damping $\Gamma$ in the superconductors is 
negligible). This term also leads to the so called interference current in 
Josephson tunnel junctions. In our case this term arises due to the proximity 
effect inducing a condensate in the normal film. In Fig.2(b) we plot the 
dependence of the zero-temperature conductance  
$S$ verses normalised voltage, $S= G_n(dI/dV)$, where $G_n$ 
is the normal state conductance. The parameter $\epsilon_{Th} 
/\epsilon_g$, the ratio of the S/N interface resistance to the N 
film resistance is chosen large so that the resistance of the system is 
determined by the interface resistance. If the damping rate $\gamma$ is 
large in comparison with $\epsilon_g$, then the DOS in the normal film is 
only weakly disturbed by the proximity effect and the peak arising in the 
subgap conductance is located at zero voltage. The S/N interface 
resistance dominates in this case, and the subgap conductance is caused by the 
interference current (the second term in Eq.(\ref{eq16})), or by 
Andreev reflection processes. In the opposite case of a small damping 
($\gamma << \epsilon_g$) or a large N/N' interface resistance a peak in the DOS 
of the N film appears in the voltage dependence of the subgap conductance 
(curves for $\gamma=0.1$ and $\gamma=0.5$). The peak in $S(V)$ is caused by a peculiarity in the 
 quasiparticle current at $eV=\epsilon_g$ (the second term in Eq.(\ref{eq16})). 

It is interesting to note that the minigap in the N film $\epsilon_g$ 
may be modulated by an 
external magnetic field if two superconductors (instead of one) are used in the 
structure
(see for example \cite{r14}), then $\epsilon_g$ depends on the phase difference 
between the superconductors $\phi$, $\epsilon_g=\epsilon_{g0}|\cos \phi|$. 
We see from Fig.2(b) that increasing $\gamma$ with respect to 
$\epsilon_g$ may lead to an increase in the conductance for some voltages. 
This means that an applied magnetic field H ($\phi \sim H$) may cause an 
increase in the conductance. This effect was predicted for a two-barrier 
structure in refs \cite{r14,r15} and observed experimentally \cite{r25,r26}.
 Note that the subgap conductance was 
first observed in a SISm system (Sm is a heavily doped semiconductor) by 
Kastalsky and Kleinsasser \cite{r31} and later by others \cite{r32,r33,r34}, 
and theoretically studied in many works \cite{r9,r36,r37,r38,r39}.

\smallskip
 {\bf 3. Nonmonotonic temperature and voltage dependence of the conductance.}
\smallskip

 In this section we analyse the voltage and temperature 
dependence of the conductance for the structure shown in Fig.1(b) 
(Andreev interferometer).  
Since the late 1970's it has been known that the conductance ($G$) of S/N 
mesoscopic structures depends on temperature ($T$) (and voltage ($V$)) in a 
non-monotonic way (see reviews \cite{r2,r3}). This behaviour was first predicted 
in Ref. \cite{r16} where a simple point S/N contact was analysed. The authors of 
Ref. \cite{r16}, using a microscopic theory and assuming that the 
energy gap in the superconductor ($\Delta$) is much less than the Thouless energy 
$\epsilon_{Th} \equiv \hbar D/L^{2}$,  
showed that the zero-bias conductance $G$ coincides at zero temperature with 
its normal state value ($G_{n}$). With increasing $T$, $G$ exhibits a 
non-monotonic behaviour, 
increasing to a maximum of $G_{max} \approx 1.25 G_{n}$ at $T_m \approx \Delta(T_m)$ 
and then decreasing to $G_{n}$ for $T>T_m$.

 Recently mesoscopic S/N structures have been fabricated in which the limit 
$\Delta >> \epsilon_{Th}$ is realised. In this case Nazarov and 
Stoof \cite{r17} (also see \cite{r18,r19,r20}) argued that the temperature dependence of 
the conductance $G$ has a similar non-monotonic behaviour with a maximum at a 
temperature comparable with the Thouless energy, while simultaniously Volkov, 
Allsopp and Lambert \cite{r21} predicted that the voltage dependence of the conductance 
in an S/N  mesoscopic structure (Andreev interferometer) has a similar form 
with a maximum at $eV_{m} \approx \epsilon_{Th}$. This non-monotonic behaviour  
has been observed both in very short S/N contacts 
\cite{r22} and in longer mesoscopic S/N structures \cite{r23,r24,r25,r27}.

 Considering the system shown in Fig.1(b) and assuming
 a weak proximity effect ($|F^{R(A)}|<<1$),  
Eq.(\ref{eq12}) and $<M^{-1}>$ can be presented in the form, 
$<M^{-1}>=1-<m>$ \cite{r20}, where 

\begin{equation}
 \left\langle m \right\rangle =\frac{1}{8}
Tr\left\langle\left(\hat{F}^{R}\right)^2  +
\left(\hat{F}^{A}\right)^2 - 2\hat{F}^{R}\hat{F}^{A}\right\rangle
\label{eq17}
\end{equation}

 The first two terms in Eq.(\ref{eq17}) determine a change in the DOS of the 
N film due to the proximity effect. This can be seen from Eq.(\ref{eq14}) by taking 
into account the smallness of $F^{R(A)}$ and the 
normalisation condition \cite{r4}

\begin{equation}
({G}^{R(A)})^{2}\hat{1} +(\hat{F}^{R(A)})^{2}=\hat{1}.
 \label{eq18}
\end{equation}

 The DOS in the N film is reduced at small energies by the proximity effect and 
therefore these terms reduce the conductance of the system. The last (anomalous) 
term in Eq.(\ref{eq17}) is analogous to the so called Maki-Thompson term in the 
paraconductivity and increases the conductance (see Fig.3). At zero 
energy both contributions cancel each other. This is seen from Eq.(\ref{eq12}). 
At $\epsilon=0$ the equations for $u^{R(A)}$ coincide with each other and 
$F^R=F^A$, hence $<m>=0$ (see Eq.(\ref{eq17})). In order to calculate the 
variation of the normalised conductance $\delta S$, one has to solve Eq.(\ref{eq12}) 
and to find the functions  $F^{R(A)}(x)$. In the case of a weak proximity effect 
Eq.(\ref{eq12}) can be linearised, and a solution is readily found,

\begin{equation} 
\hat{F}^{R(A)}(\epsilon,x)=i\hat{\sigma}_y r F^{R(A)}_S \left[
\sinh (k^{R(A)}(L-x)) \right] \cos (\phi /2)/(\theta \cosh \theta)^2
 \label{eq19}
\end{equation}

 where $r$ is the ratio of the N film resistance and the S/N interface 
resistance in the normal state, $\phi$ is the phase difference between 
the superconductors, for simplicity we assume that $L_1 << L$. 
Knowing $\hat{F}^{R(A)}$ we can easily calculate 
$<m>$ and $\delta S$ using Eqs.(\ref{eq10}-\ref{eq11}). The function 
$<m>$ which determines the voltage dependence of the zero-temperature 
conductance is equal to,

\begin{equation} 
<m> =\frac{r^2}{8} \left[ Re\left(\frac{\sinh2\theta}{2\theta} -1 \right)
/(\theta \cosh \theta)^2 - \left(\frac{\sinh2\theta_1}{2\theta_1}-
\frac{\sin2\theta_2}{2\theta_2}\right)
/|\theta \cosh \theta|^2 \right](1+\cos\phi)
 \label{eq20}
\end{equation}

 where $\theta=\theta_1+i\theta_2$. The first term in Eq.(\ref{eq20})  
is related to a change in the DOS and the second term is due to the anomalous 
term $<F^RF^A>$. In Fig.3 we present the total conductance variation $\delta S$ 
vs the normalised voltage $V$ (in units of the Thouless energy) at zero 
temperature, the contribution $\delta S_{DOS}$ due to a 
change of the DOS, and the anomalous term $\delta S_{an}$ for $\phi=0$. 
The variation $\delta S$ has a maximum at $eV \approx \epsilon_{Th}$ \cite{r21}. 
The temperature dependence $\delta S(T)$ found with the help of Eqs.
(\ref{eq10}-\ref{eq11}) has a similar form provided $T<<\Delta,\epsilon_{Th}
<<\Delta$  \cite{r17,r18,r19,r20}. It is interesting to note that as $T$ increases  
 $\delta S_{an}$ decays slowly (non-exponentially) as ($\epsilon_{Th}/T$) at
$T>>\epsilon_{Th}$. This law leads to long-range, phase-coherent effects in 
mesoscopic structures \cite{r20,r28,r29,r31}. One can see from Eq.(\ref{eq19}) that the 
functions $|F^{R(A)}|$ have a peak at $\epsilon\approx\Delta$ because the 
functions $F^{R(A)}_S$ have a singularity at the energy $|\epsilon|=\Delta$ 
(see Eq.(\ref{eq3})). With increasing temperature $T$ the order parameter 
$\Delta(T)$ starts decreasing and near $T_c$ may be comparable with $\epsilon_{Th}$, 
hence the function $<m>$ and $\delta S$ may have one more maximum. In Fig.4 
the temperature dependence of $\delta S$ is presented over a wide temperature 
range ($0<T<T_c$)  for a structure similar to that shown in Fig.1b (where 
$L_1=0$, i.e. only one superconductor is in contact with the 
normal film). We see that besides the main, low-temperature maximum in 
 $\delta S(T)$ there is an additional maxinum near  $T_c$. The magnitude 
of both maxima depends strongly on the depairing rate $\gamma$ in the N 
film (in order to account for $\gamma$, the energy $\epsilon$ must be replaced 
with $\epsilon \pm i \gamma $) \cite{r35}.

\smallskip
 {\bf 4. Negative Josephson current 
in a 4 terminal S/N/S structure.}
\smallskip

 In the preceding section we calculated the conductance $S$ of the 
Andreev interferometer (see Fig.1b), determined by the nonequilibrium  
distribution function $f$ (see Eq.(\ref{eq2})) which varies in space. 
In the limit of a weak proximity effect we obtain from 
Eq.(\ref{eq9})

\begin{equation}
f(\epsilon,x) = J(\epsilon)[x +  \int^{x}_{0}
dx_1 m(x_1) ]
\label{eq21}
\end{equation}

Here $J(\epsilon)$ is a function even in energy $\epsilon$ and defined in Eq.(10).
Therefore $f(\epsilon,x)$ increases almost linearly from zero (at $x = 0$) to 
$F_N$ (at $x = L$). The function $<m>$ determined by the condensate functions 
depends on the phase difference $\phi$ (see Eq.(20)), increasing $\phi$ 
(for example, in an applied magnetic field) causes the function $<m>$ and the 
conductance $S$ to oscillate. This oscillatory behaviour of the 
conductance has been observed in many experiments \cite{r23,r24,r26,r32,r33,r41} starting from the 
pioneering work \cite{r40} and later explained theoretically \cite{r14,r17,r42}.

 One can formulate the inverse problem, how the non-equilibrium distribution 
function affects, for example, the condensate current, which is 
determined by the expression,

\begin{equation}
I_s=\frac{\sigma d}{8} Tr \left[ \hat{\sigma}_z \int d\epsilon \left[
\hat{F}^R \partial\hat{F}^R -\hat{F}^A \partial\hat{F}^A \right]f_o
\right]
\label{eq22}
\end{equation}

where $f_o$ is a function odd in energy $\epsilon$ and determined in 
Eq.(\ref{eq5}). By taking the trace of this equation in the limit of a 
weak proximity effect we obtain,

\begin{equation}
\partial_x [(1-m_o)\partial_xf_o]=0
\label{eq23}
\end{equation}

 here $m_o = (1/8) Tr (\hat{F}^R +\hat{F}^A)^2$. The current $I_s$ is of the 
order of the small parameter $r^2$, therefore in finding $f_o$ we can neglect 
$m_o$ and obtain that 
$f_o$ is constant in space and equal to $f_o(\epsilon)=F_{N0}$ 
(see Eq.(\ref{eq7})). The function $F_{N0}(\epsilon)$ is an equilibrium distribution 
function (odd in $\epsilon$) in the N reservoirs at a given potential $V$,
and is shifted with respect to the function $F_{0S}= \tanh \epsilon\beta$ by 
the value $V$ (the potential in the superconductors is set equal to zero). 
The function $F_{N0}$ is an even function of $V$ and therefore does not 
depend on whether  the additional (control) current flows from the N 
reservoirs to the superconductors (the potential $V$ has the same sign in both 
reservoirs) or from one N reservoir to another one ($V$ has opposite signs in 
the reservoirs). A relative shift of the distribution functions in N and S 
controlled by the applied voltage $V$ leads to a dependence of the 
Josephson critical current $I_c$ on $V$. In order to determine this 
dependence, we need to find the condensate functions $\hat{F}^{R(A)}$, 
obeying the linearised Usadel equation which in this case has the form,

\begin{equation}
\partial_{xx} \hat{F}^{R(A)} - \left(k^{R(A)}\right)^2 \hat{F}^{R(A)}=
-k_{b}^{2} w \hat{F}^{R(A)}_S 
\left[\delta\left(x+L_{1} \right)+\delta\left(x-L_{1} \right)
\right]
\label{eq24}
\end{equation}

 the functions $\hat{F}^{R(A)}_S$ are given by Eq.(\ref{eq3}). Solving Eq.(\ref{eq24}) 
and substituting the solutions into Eq.(\ref{eq22}), we find $I_s =I_c \sin \phi$, where 

\begin{equation}
I_c(V) R_L=(\pi T)r^2 \Re \left[ \sum_{n=0}^{\infty} \frac{\Delta^2}{\Delta^2+(\omega_n+ieV)^2}
\frac{ \sinh ^2 (\theta_n (L-L_1)/L)}{\theta_n \sinh (2 \theta_n)} \right]
\label{eq25}
\end{equation}

 here $\theta_n = L [(2(\omega_n+ieV)+\gamma)/D]^{1/2}$, $ \omega_n= \pi T (2n+1)$ 
is the Matsubara frequency. In Fig.5 we show the dependence of the 
Josephson critical current $I_c$ on $V$. We see that $I_c$ changes sign 
when $eV \approx \epsilon_{Th}$. The $I_c$ sign reversal effect 
($\pi$ - contact) was first predicted in 
Ref \cite{r44} where a S/M/S contact was considered (M is an insulating layer 
with magnetic impurites). The same effect was analysed later in Ref \cite{r45} 
where a S/F/S contact was considered. The possibility to change of the 
$I_c$ sign by a control voltage (or current) in a 4-terminal  S/N/S 
diffusive structure was first studied in Ref \cite{r46}  for the 1- dimentional case 
of a short spacing between superconductors ($L_1<\hspace*{-0.4cm}_{\sim} 
\xi_S$) and in Refs.\cite{r20,r29,r48,r49} for the case of a longer spacing.
The physical reason for the sign reversal of $I_c$ in 4 terminal S/N/S 
structures is a difference in the distribution functions $f_o$ in the 
N film and in the superconductors. These functions have an equilibrium 
form corresponding to the electric potential $V$ and zero respectively. 
The 1-dimensional case was analysed in \cite{r20,r29,r46,r48}. In 
Ref \cite{r49} the 2-d case was 
analysed, however in finding the distribution funtion $f_o$ and 
the condensate functions  $F^{R(A)}$ the authors considreed 
opposite limits, $d_s<<d$ and $d_s>>d$, where $d_s,d$
 are the width of the S and N films respectively. Therefore
the region of applicability of the results obtained in Ref \cite{r49} 
remains unclear. One 
can show that qualitatively Eq.(\ref{eq25}) in limit $d_s << d$ remains valid 
for the 2-d case, 
i.e. $I_cR_N$ is proportinal to the small parameter $r^2$ (r is the ratio of the 
N film resistance to the resistance of the S/N contact).  Similar effects in 
ballistic systems were studied in Refs.\cite{j1,j2,j3}.

 Note, a formal analogy between the two possibilities of the $I_c$ sign 
reversal effect: a structure with magnetic material and a shift of the 
distribution functions in a 4-terminal S/N/S structure. In the first case 
the Green's functions $F^{R(A)}
(\epsilon)$ in the F layer differ from those in the N layer by a shift 
in energy $\epsilon \rightarrow \epsilon+h$, where h is the exchange energy 
\cite{r45}. The distribution function $f_o$ in the F layer coincides with the 
distribution function in the superconductor $F_{0S}=\tanh \epsilon \beta$. 
Meanwhile in the 4-terminal S/N/S contact the distribution function in the 
N layer is shifted with respect to $F_{S0}$, $f_o(\epsilon)=[F_{0S}(\epsilon+eV)
+ F_{0S}(\epsilon-eV)]/2$. Making the substitution $\epsilon \rightarrow \epsilon
\pm eV$ in the integral in Eq.(\ref{eq25}), one can reduce the case of a 
4-terminal S/N/S contact to the case of a S/F/S contact with $h=eV$. The sign
reverse of $I_c$ in S/F/S contacts has, presumably, not yet been observed. On 
the other hand in a recent publication \cite{r50} experimental evidence for 
the change of the sign of $I_c$ in 4-terminal S/N/S contacts was obtained.

\smallskip
 {\bf 5. Conclusions.}
\smallskip

 The quasiclassical Green's function technique we have used complements the 
approach based on the scattering matrix method and the Landauer formula 
for the conductance. This technique allows one to describe the transport 
properites of S/N mesoscopic structures over a wide range of parameters. 
With the help of this approach the conductance of a double-barrier S/N/N' 
structure was calculated. We have shown that in the differential subgap 
conductance $S$ has a peak situated at zero or finite bias voltage 
(depending on the relation between the S/N and N/N' interface resistances). 
When the temperature is lowered below $T_c$, the variation of the 
conductance may be both positive and negative. Therefore the conductance 
decreases or increases with increasing an external magnetic field.

 The conductance in S/N structures of the types of the Andreev interferometer
 or S/N point contact depends on T or V in a non-monotonic way. As a 
function  of temperature T the zero-bias conductance variation $\delta S$ 
increases from zero, has a maximum at $T_1$ of order the Thouless energy 
$\epsilon_{Th}$ (if $\epsilon_{Th}<<\Delta$) and decays to zero at $T \rightarrow 
T_c$. Near $T_c$ a second smaller peak in  $\delta S$ may occur at a temperature 
$T_2$ corresponding to the condition $\Delta(T_2) \approx \epsilon_{Th}$. 

 An interesting effect may occur in 4 - terminal S/N/S mesoscopic structures 
(see for example Fig.1b). If an additional (control) current flows through the N film, 
then the 
distribution functions in the N and S films are different. Due to this difference 
the critical Josephson current may change sign and the S/N/S contact may become a 
$\pi$-contact. Most of the effects discussed in this paper have been observed 
experimentally, although some predictions such as the appearance of the second maximum near 
$T_c$ in $\delta S$ have not yet been studied experimentally.

  A. F. Volkov is grateful to the Royal Society, to the 
Russian grant on superconductivity (Project 96053) 
and to CRDF (project RP1-165) for financial support.


\begin{figure}
\centerline{\psfig{figure=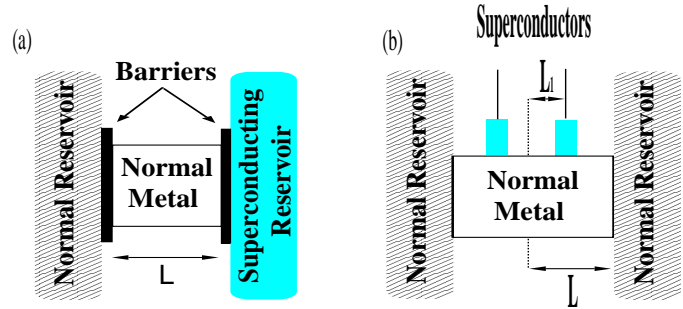,width=10cm,height=6cm}}
\caption{The structures considered.}
\label{fig1}
\end{figure}

\begin{figure}
\centerline{\psfig{figure=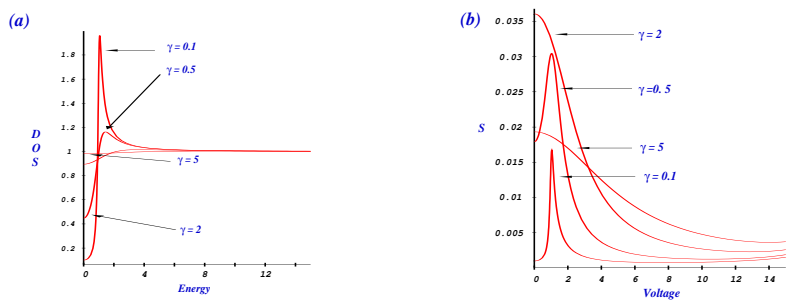,width=12cm,height=4cm}}
\caption{(a) The energy dependence of the DOS for 
the structure in Fig.1a, for different $\gamma$; here $\Delta=20,
\Gamma=0.1$ and $\epsilon_d=10$. (b) The zero temperature conductance $S$ 
(Eq.(\ref{eq15})), normalised to the conductance of the normal metal in the 
norml state, dependence on voltage, using the same parameters as for Fig.2a.
 All energies are measured in units of $\epsilon_{Th}$.}
\label{fig2}
\end{figure}

\begin{figure}
\centerline{\psfig{figure=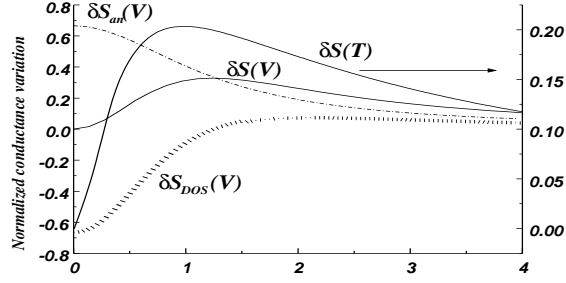,width=8cm,height=6cm}}
\caption{Normalized conductance $\delta S$ vs. normalized voltage
    $eV/\epsilon _{L} $ at $T=0$ and vs. normalized temperature
    $T/\epsilon _{L} $ at $V=0$ for the structure shown in Fig.1b ($L_1=0$).}
\label{fig3}
\end{figure}

\begin{figure}
\centerline{\psfig{figure=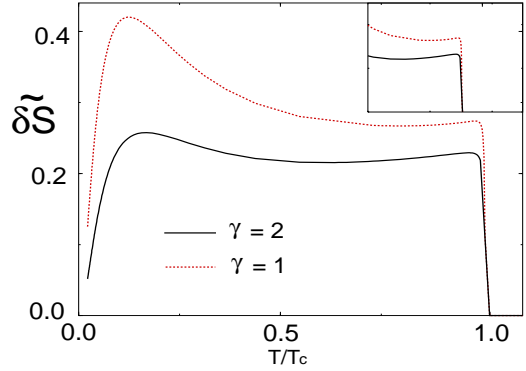,width=7cm,height=5cm}}
\caption{Dependence of $\delta \tilde{S}$ on temperature, for the cross
 geomentry in the 
weak proximity limit, for different depairing rates $\gamma$. With  
$\Delta_{0}=20$,$\Gamma=0.1$,$L_1/L=0$. The inset is an enlargement of the 
second maximum around $T/T_c=1.0$. Note the $\delta 
\tilde{S}=\delta S(2r_b)^2$.}
\label{fig4}
\end{figure}

\begin{figure}
\centerline{\psfig{figure=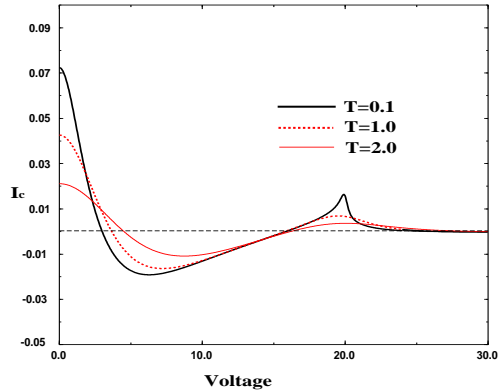,width=7cm,height=7cm}}
\caption{Critical current $I_c$ dependence on voltage for the 4 terminal 
Andreev interferometer, for several temperatures ($T$). The critical current is 
measured in units $\epsilon_{Th}/(eR_Lr^2)$
 and $T,V$ in $\epsilon_{Th}$;$\Delta=20,\gamma=0.2, 
L_1/L=0.5$. }
\label{fig5}
\end{figure}

\end{document}